\documentclass[iop]{emulateapj}

\usepackage{atbegshi}
\def\hackaltaffiltext#1#2{\AtBeginShipoutNext{\footnotetext[#1]{#2}\stepcounter{footnote}}}

\def\ltsima{$\; \buildrel < \over \sim \;$}
\def\simlt{\lower.5ex\hbox{\ltsima}}
\def\gtsima{$\; \buildrel > \over \sim \;$}
\def\simgt{\lower.5ex\hbox{\gtsima}}
\def\kpc{{\rm\,kpc}}

\def\msun{{\rm\,M_\odot}}
\def\lsun{{\rm\,L_\odot}}

\def\pc{{\rm\,pc}}

\def\deg{^\circ}

\def\s{\ifmmode \widetilde \else \~\fi}
\def\={\overline}

\def\spose#1{\hbox to 0pt{#1\hss}}

\def\lta{\mathrel{\spose{\lower 3pt\hbox{$\mathchar"218$}}
     \raise 2.0pt\hbox{$\mathchar"13C$}}}
\def\gta{\mathrel{\spose{\lower 3pt\hbox{$\mathchar"218$}}
     \raise 2.0pt\hbox{$\mathchar"13E$}}}
\def\Dt{\spose{\raise 1.5ex\hbox{\hskip3pt$\mathchar"201$}}}    
\def\dt{\spose{\raise 1.0ex\hbox{\hskip2pt$\mathchar"201$}}}    

\def\dotsfill{\leaders\hbox to 1em{\hss.\hss}\hfill}

\def\Gyr{{\rm\,Gyr}}
\def\FeH{{\rm[Fe/H]}}

\shorttitle{SMASH~1}
\shortauthors{N. F. Martin et al.}

\begin{document}


\title{SMASH 1: a very faint globular cluster disrupting in the outer reaches of the LMC?}


\author{Nicolas F. Martin$^{1,2}$, Valentin Jungbluth$^1$, David L. Nidever$^{3,4}$, Eric F. Bell$^5$, Gurtina Besla$^6$, Robert D. Blum$^7$, Maria-Rosa L. Cioni$^{8,9,10}$, Blair C. Conn$^{11,12}$, Catherine C. Kaleida$^{13}$, Carme Gallart$^{14,15}$, Shoko Jin$^{16,17}$, Steven R. Majewski$^{18}$, David Martinez-Delgado$^{19}$, Antonela Monachesi$^{20}$, Ricardo R. Mu\~noz$^{21}$, Noelia E. D. No\"el$^{22}$, Knut Olsen$^7$, Edward W. Olszewski$^4$, Guy S. Stringfellow$^{23}$, Roeland P. van der Marel$^{24}$, A. Katherina Vivas$^{13}$, Alistair R. Walker$^{13}$, Dennis Zaritsky$^{4}$} 

\email{nicolas.martin@astro.unistra.fr}

\altaffiltext{1}{Observatoire astronomique de Strasbourg, Universit\'e de Strasbourg, CNRS, UMR 7550, 11 rue de l'Universit\'e, F-67000 Strasbourg, France}
\altaffiltext{2}{Max-Planck-Institut f\"ur Astronomie, K\"onigstuhl 17, D-69117 Heidelberg, Germany}
\altaffiltext{3}{Large Synoptic Survey Telescope, 950 North Cherry Ave, Tucson, AZ 85719}
\altaffiltext{4}{University of Arizona, Steward Observatory, 933 North Cherry Ave, Tucson, AZ 85719}
\altaffiltext{5}{Department of Astronomy, University of Michigan, 1085 S. University Ave., Ann Arbor, MI 48109-1107, USA}
\altaffiltext{6}{Department of Astronomy, University of Arizona, AZ 85721-0065, USA}
\altaffiltext{7}{National Optical Astronomy Observatory, 950 N Cherry Ave, Tucson, AZ 85719, USA} 
\altaffiltext{8}{Universit\"{a}t Potsdam, Institut f\"{u}r Physik und Astronomie, Karl-Liebknecht-Str. 24/25, 14476 Potsdam, Germany}
\altaffiltext{9}{Leibniz-Institut f\"{u}r Astrophysik Potsdam (AIP), An der Sternwarte 16, 14482 Potsdam, Germany}
\altaffiltext{10}{University of Hertfordshire, Physics Astronomy and Mathematics, Hatfield AL10 9AB, United Kingdom}
\altaffiltext{11}{Research School of Astronomy and Astrophysics, The Australian National University, Mt Stromlo Observatory, via Cotter Road, Weston, ACT 2611, Australia}
\altaffiltext{12}{Gemini Observatory, Recinto AURA, Colina El Pino s/n, La Serena, Chile}
\altaffiltext{13}{Cerro Tololo Inter-American Observatory, National Optical Astronomy Observatory, Casilla 603, La Serena, Chile}
\altaffiltext{14}{Instituto de Astrof\'{i}sica de Canarias, La Laguna, Tenerife, Spain}
\altaffiltext{15}{Departamento de Astrof\'{i}sica, Universidad de La Laguna, Tenerife, Spain}
\altaffiltext{16}{Kapteyn Astronomical Institute, University of Groningen, PO Box 800 9700AD Groningen, The Netherlands}
\altaffiltext{17}{Leiden Observatory, Leiden University, PO Box 9513, NL-2300 RA Leiden, The Netherlands}
\hackaltaffiltext{18}{Department of Astronomy, University of Virginia, Charlottesville, VA 22904, USA}
\hackaltaffiltext{19}{Astronomisches Rechen-Institut, Zentrum f\"ur Astronomie der Universit\"at Heidelberg,  M{\"o}nchhofstr. 12-14, 69120 Heidelberg, Germany}
\hackaltaffiltext{20}{Max-Planck-Institut f\"ur Astrophysik, Karl-Schwarzschild-Str. 1, D-85748 Garching, Germany}
\hackaltaffiltext{21}{Departamento de Astronom\'ia, Universidad de Chile, Camino del Observatorio 1515, Las Condes, Santiago, Chile}
\hackaltaffiltext{22}{Department of Physics, University of Surrey, Guildford, GU2 7XH, UK}
\hackaltaffiltext{23}{Center for Astrophysics and Space Astronomy, University of Colorado, 389 UCB, Boulder, CO, 80309-0389, USA}
\hackaltaffiltext{24}{Space Telescope Science Institute, 3700 San Martin Drive, Baltimore, MD 21218}

\begin{abstract}
We present the discovery of a very faint stellar system, SMASH 1, that is potentially a satellite of the Large Magellanic Cloud. Found within the Survey of the \textsc{Ma}gellanic Stellar History (SMASH), SMASH 1 is a compact ($r_h = 9.1^{+5.9}_{-3.4}\pc$) and very low luminosity ($M_V = -1.0\pm 0.9$, $L_V=10^{2.3\pm 0.4}\lsun$) stellar system that is revealed by its sparsely populated main sequence and a handful of red-giant-branch candidate member stars. The photometric properties of these stars are compatible with a metal-poor ($\FeH=-2.2$) and old (13 \Gyr) isochrone located at a distance modulus of $\sim18.8$, i.e. a distance of $\sim57\kpc$. Situated at $11.3\deg$ from the LMC in projection, its 3-dimensional distance from the Cloud is $\sim13\kpc$, consistent with a connection to the LMC, whose tidal radius is at least $16\kpc$. Although the nature of SMASH~1 remains uncertain, its compactness favors it being a stellar cluster and hence dark-matter free. If this is the case, its dynamical tidal radius is only $\lta19\pc$ at this distance from the LMC, and smaller than the system's extent on the sky. Its low luminosity and apparent high ellipticity ($\epsilon=0.62^{+0.17}_{-0.21}$) with its major axis pointing toward the LMC may well be the tell-tale sign of its imminent tidal demise. 
\end{abstract}


\section{Introduction}
The commissioning of the Dark Energy Camera (DECam) mounted in the CTIO Blanco 4-meter telescope has triggered numerous discoveries of previously unknown nearby stellar systems, most of which are thought to be satellites of the Milky Way or of the Magellanic Clouds. The Dark Energy Survey (DES) itself has enabled the discovery of more than a dozen new candidate dwarf galaxies and globular clusters \citep{bechtol15,drlica-wagner15,kim15b,koposov15,luque16}. Other complementary surveys, such as the Stromlo Milky Way Satellite (SMS) survey and the Survey of the \textsc{Ma}gellanic Stellar History (SMASH) have further revealed other (very) faint satellites \citep{kim15a,martin15,kim16}. All the new discoveries share a similarly low surface brightness ($\mu\gta26$mag/arcsec$^2$) that explains why they went unnoticed in previous photographic plate surveys.

Some of the new systems are clearly unrelated to the Magellanic Clouds but the discovery of so many new satellites in the physical vicinity of the LMC and SMC naturally leads to the conclusion that a significant fraction of the new discoveries were brought into the Milky Way halo by this infalling group \citep{drlica-wagner15,koposov15,martin15}. They can, in turn, be used to provide unique constraints on the accretion timing of the Magellanic group and its properties \citep{deason15,jethwa16}.

The properties of the individual satellites are also interesting in their own right as they could hold important clues on the group preprocessing that most dwarf galaxies are thought to have experienced before being accreted onto a more massive host \citep[e.g.,][for the mass-scale relevant here]{deason14,wetzel15}. Work has only just started to better characterize the new systems \citep[e.g.,][]{kirby15a,simon15,walker15,walker15b} and, in particular, to understand which are dark-matter dominated dwarf galaxies, which are globular clusters, and how the dynamical LMC-SMC group environment has affected them.

\section{The SMASH data and discovery}
SMASH is a NOAO community survey conducted with DECam on the CTIO Blanco 4m with $\sim40$ nights to gather deep photometry  over $\sim$2,400 deg$^2$ of the Magellanic system at 20\% filling factor. The main goal of the survey is the study of the complex stellar structures of the Magellanic system (the Clouds themselves, the Magellanic Bridge and the leading part of the Magellanic Stream; D. Nidever et al., in preparation).

The DECam Community Pipeline \citep{valdes14} produces InstCal image data products (calibrated, single-frame reduced image with instrument signature removed, WCS and rough photometric calibrations applied), which we accessed through the NOAO Science Archive Server\footnote{\url{https://www.portal-nvo.noao.edu}}. We then use the PHOTRED\footnote{\url{https://github.com/dnidever/PHOTRED}} pipeline \citep{nidever11} to perform the rest of the photometric reduction. PHOTRED is an automated PSF photometry pipeline based on DAOPHOT \citep{stetson87,stetson94} and SExtractor \citep{bertin96}. It performs WCS fitting, single-image PSF photometry (ALLSTAR), forced PSF photometry of multiple images with a master source list created from a deep stack of all exposures (ALLFRAME), and aperture correction. The data used here are then photometrically calibrated as follows. First, a relative photometric calibration is performed using an `\"ubercal' technique and overlapping exposures. Next, APASS\footnote{\url{https://www.aavso.org/apass}} is used to apply a single, absolute photometric calibration per field for the $g$- and $i$-band, and the position of the stellar locus is used to roughly calibrate the $u$-, $r$- and $z$-band.

All magnitudes are de-reddened following \citet[][assuming $R_V=3.1$]{schlafly11}, and denoted with the ``0'' subscript. Finally, we select only star-like objects by enforcing a cut on the DAOPHOT sharpness and $\chi^2$ parameters ($|\textrm{sharp}|<0.5$, $\chi^2<1.1$). We further use the SExtractor CLASS\_STAR stellar probability index to remove objects that are clearly galaxies ($\textrm{prob}<0.01$). Most of the culling is produced by the sharpness cut.

SMASH~1 was discovered through a visual inspection of the stellar distribution of stars that could correspond to red giant branch or main sequence stars ($-0.1\lta(g-r)_0\lta 1.0$; \citealt[e.g.,][]{koposov08}) for the 100 SMASH fields with reduced data available as of June 2016. This investigation revealed a single stellar overdensity, SMASH~1, and this letter therefore uses only data from a single field observed during the night of UT 2014 January 6 and centered on $(\alpha,\delta)=(6\mathrm{h}25\mathrm{m}24.4\mathrm{s},-80\deg00'13\farcs9)$.

\section{Properties of SMASH~1}
\begin{figure*}
\begin{center}
\includegraphics[width=0.33\hsize,angle=270]{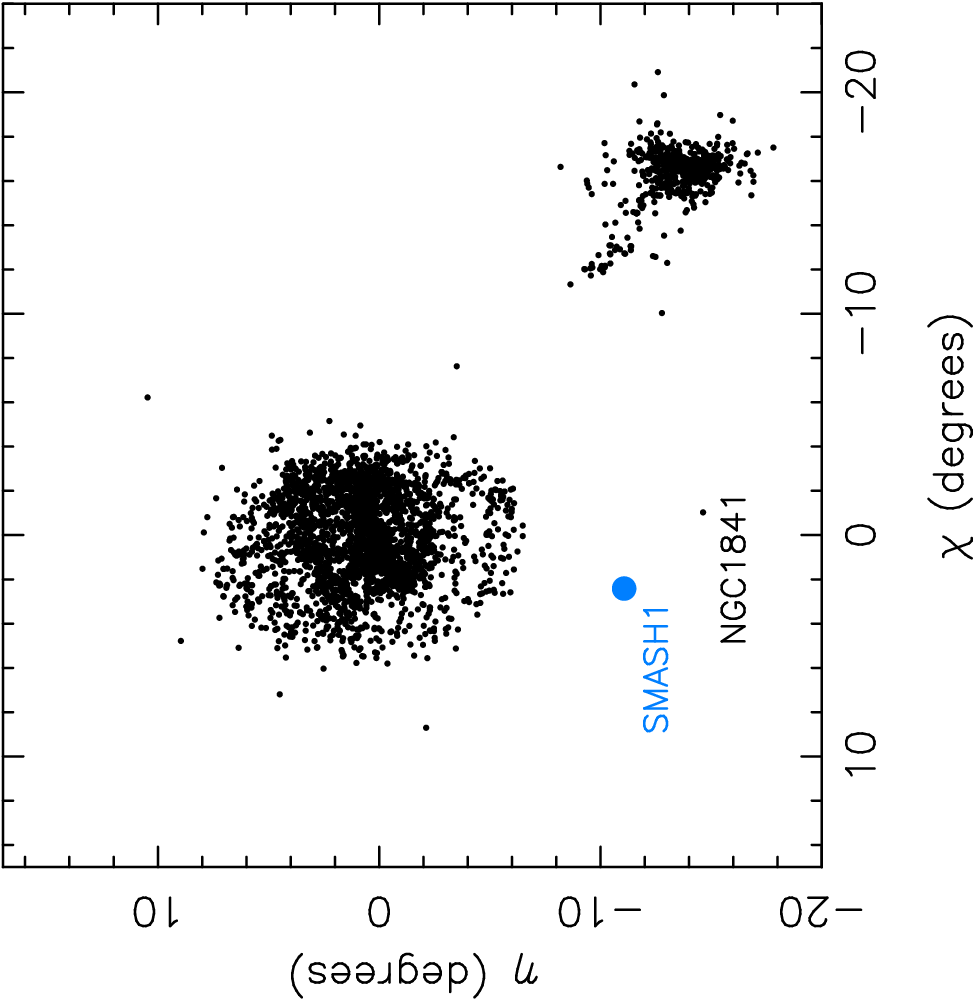}
\hspace{1cm}
\includegraphics[width=0.33\hsize,angle=270]{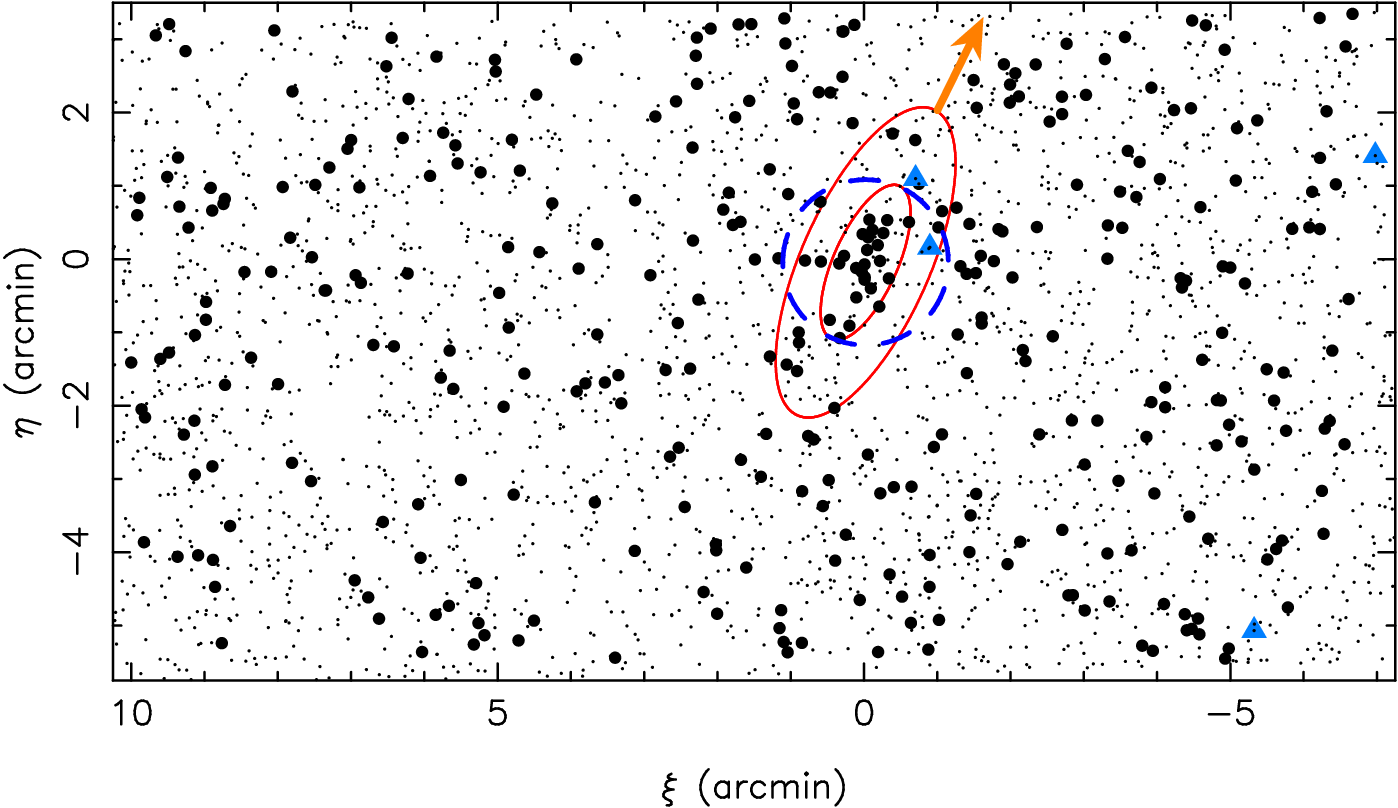}
\caption{\label{map_Cl}\label{map}Left: Distribution of clusters around the LMC and the SMC from \citet{bica08} and \citet{pieres16}. SMASH~1 is represented by the large blue dot, at a distance from the LMC where only a few clusters are known, including the old and metal-poor NGC~1841. Right: Distribution of stellar sources on chip 47 of the field that contains SMASH~1 (small dots), with stars compatible with the main sequence of SMASH~1 highlighted as large black dots. Four potential SMASH~1 horizontal branch stars are displayed as blue triangles. The red ellipses represent the 2- and 4-half-light-radius regions around the stellar system while the dashed blue circle marks the upper limit on the dynamical tidal radius of the system at its current location under the assumption that it is a purely stellar system. The tangential-projection coordinates are centered on SMASH~1. The orange arrow points towards the LMC centroid.}
\end{center}
\end{figure*}

\begin{figure*}
\begin{center}
\includegraphics[width=0.375\hsize,angle=270]{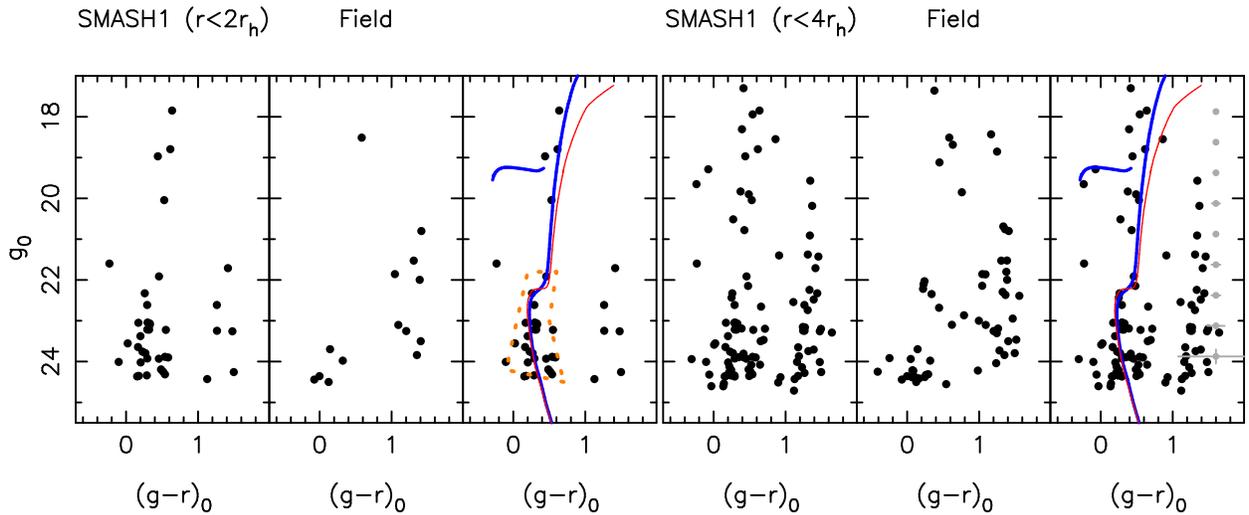}
\caption{\label{CMDs}Left: CMD of stellar-like objects within 2 half-light radii of SMASH~1's centroid ($1\farcm1$ along the major axis) and that of a comparison field with the same area. The CMD of SMASH~1 is repeated in the third panel, but this time with the $13\Gyr$ and $\FeH=-2.2$ \textsc{Parsec} isochrone \citep{bressan12} at a distance modulus of 18.8 overlaid in blue. The thin red isochrone corresponds to a younger ($8\Gyr$), $\FeH=-1.4$ isochrone at a distance modulus of 19.0 that also tracks the main sequence turnoff of SMASH1. The dotted-line, orange polygon corresponds to the selection box used to infer the structural properties of SMASH~1. Right: Same plots for the region within 4 half-light radii of SMASH~1's centroid ($2\farcm2$ along the major axis). The gray points with error bars in the left-most panel show the average photometric uncertainties.}
\end{center}
\end{figure*}

SMASH~1 is found in the outskirts of the LMC, in a region that nevertheless hosts a few LMC/SMC clusters (left panel of Figure~\ref{map}). Figure~\ref{map} (right panel) presents the distribution of all star-like sources on chip 47 (CCDNUM=47) as small dots and highlights as large dots objects selected within a color-magnitude diagram (CMD) selection box tailored to isolate the main sequence stars of SMASH~1. The stellar system corresponds to a significant ($\sim4\sigma$) overdensity compared to the field population and appears rather elongated. The CMD of stars in SMASH~1 is displayed in Figure~\ref{CMDs} for both a 2- and a 4-half-light-radius region ($2r_h$ or $4r_h$, $1\farcm1$ or $2\farcm2$ along the system's major axis; see below for the inference of the structural parameters). Compared to the field CMDs shown in the same figure, SMASH~1 is revealed by a few tens of main sequence stars with $g_0\gta22.0$ and $0.0\lta(g-r)_0\lta0.6$. In the CMD within $2r_h$, a handful of likely red giant branch (RGB) stars is also visible, aligned between $[(g-r)_0,g_0]=(0.4,22.0)$ and $(0.6,18.0)$. Finally, the $4r_h$ CMD reveals two potential horizontal-branch stars located around $[(g-r)_0,g_0]=(-0.2,19.5)$. After comparison by eye with a family of \textsc{Parsec} isochrones \citep{bressan12}, we conclude that all these features are well reproduced by the $13\Gyr$, $\FeH=-2.2$ ($Z=10^{-4}$) \textsc{Parsec} isochrone shifted to a distance modulus of $\sim18.8$ ($\sim57\kpc$). A significantly younger ($8\Gyr$) and more metal-rich isochrone ($\FeH=-1.4$) can provide a reasonable fit to the main sequence and main sequence turnoff for a distance modulus of 19.0 ($\sim64\kpc$), but it becomes too red to overlap with the likely RGB stars and is therefore disfavored (see Figure~\ref{CMDs}).

To determine the structure of SMASH~1, we apply the algorithm of \citet{martin08b}, updated in \citet{martin16c} with a full Markov Chain Monte Carlo treatment. This algorithm infers the posterior probability distribution functions (PDFs) for the 6 parameters of a family of exponential radial profiles, allowing for elliptical stellar distributions. The six parameters are: the centroid of the system, $(\alpha,\delta)$; the exponential half-light radius along the major axis, $r_h$; the ellipticity, $\epsilon$, defined as $1-b/a$ where $a$ and $b$ correspond to the major and minor axis, respectively; the position angle of the major axis East of North, $\theta$; and the number of stars, $N^*$, within the chosen CMD selection box that focuses on the main sequence of SMASH~1 (Figure~\ref{CMDs}). The model assumes a flat field contamination that is determined from the normalization of the likelihood function. Despite visible RGB candidates in Figure~\ref{CMDs}, we choose to avoid the RGB region as it is heavily contaminated by foreground stars and does not significantly help in improving the inference.

\begin{figure}
\begin{center}
\includegraphics[width=0.6\hsize,angle=270]{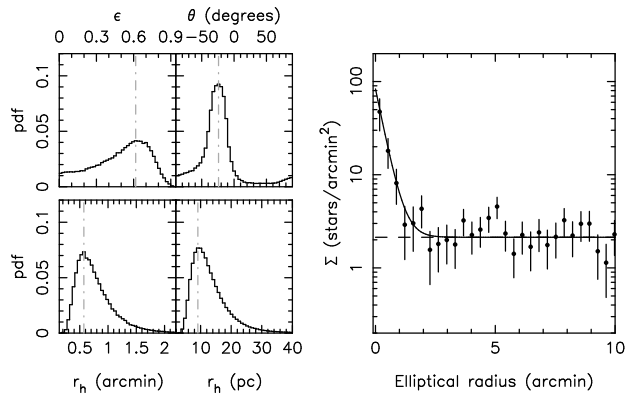}
\caption{\label{structure}Left panels: Marginalized PDFs for 3 of the 6 structural parameters of SMASH~1 (the ellipticity $\epsilon$, the position angle $\theta$, and the half-light radius $r_h$). The modes of the distributions are represented by the gray dot-dashed lines. Right panel: Radial density profile of the exponential model favored by the algorithm (full line), compared to the data within the CMD selection box (dots), binned following the favored ellipticity, position angle, and centroid. The error bars represent Poissonian uncertainties on the star counts and the dashed line shows the favored contamination level.}
\end{center}
\end{figure}

The PDFs\footnote{From these PDFs, the favored models are taken to be the modes of the distributions. Credible intervals are calculated to correspond to the highest density interval containing 68\% of the posterior PDF.} resulting from the application of the algorithm are displayed in Figure~\ref{structure}. They confirm the elongation of the system ($\epsilon=0.62^{+0.17}_{-0.21}$) and that SMASH~1 is a compact object with $r_h = 0\farcm57^{+0.32}_{-0.18}$ along the major axis. For the distance modulus of $\sim18.8$ determined above by comparison with old and metal-poor isochrones, this translates to a physical size of $9.1^{+5.9}_{-3.4}\pc$. The uncertainty on this measurement does not account for the uncertainty on the distance to the system, but this latter one is undoubtedly smaller than the $\sim45$\% uncertainty on the angular $r_h$.

The total luminosity of the system is also determined using the framework presented in \citet{martin08b} and \citet{martin16c}: the \textsc{Parsec} isochrone and luminosity function of a $13\Gyr$ old stellar population with $\FeH=-2.2$, assuming a canonical IMF \citep{kroupa01}, are placed at a distance modulus of 18.8 and convolved by the photometric uncertainties to build the CMD PDF that represents the likelihood of a SMASH~1 star in color-magnitude space. After drawing a target number of stars, $N^*_i$, from the structural parameter chain, stars are drawn from the CMD PDF and further checked against the completeness\footnote{The completeness functions are determined through artificial star tests and detailed in Nidever et al. (in prep.).} at their magnitudes until $N^*_i$ of them fall in the CMD selection box used to determine the structural parameters. Summing the flux of all stars drawn, whether in the selection box or not, yields the total luminosity of a system that has as many stars as SMASH~1 in the selection box. This procedure returns less noisy results than would otherwise be achieved by summing the flux of observed stars as these are severely contaminated with foreground stars, especially along the RGB, where fluxes are large and membership uncertain. Repeating the exercise 500 times for different random drawings of the distance modulus and $N^*_i$ further allows us to determine the uncertainties on the total luminosity of the satellite. We infer a total luminosity\footnote{Here as well, we do not formally account for the impact of the distance uncertainty on the inference of the luminosity, but it would be negligible compared to the large uncertainty coming from the `CMD shot noise.'} of $L_V=10^{2.3\pm 0.4}\lsun$ or $M_V = -1.0\pm 0.9$.

All the properties of SMASH~1 are summarized in Table~\ref{properties}.

\begin{table}
\caption{\label{properties}Properties of SMASH~1}
\begin{tabular}{c|c}
\hline
$\alpha$ (J2000) & $6^\mathrm{h}20^\mathrm{m}59.9^\mathrm{s}$\\
$\delta$ (J2000) & $-80\deg23'44\farcs7$\\
$\ell$ & $292.14\deg$\\
$b$ & $-27.99\deg$\\
$(m-M)_0$ & $\sim18.8$\\
Heliocentric Distance & $\sim57\kpc$\\
LMC Distance & $\sim13\kpc$\\
$M_{V}$ & $-1.0\pm0.9$\\
$L_{V}$ & $10^{2.3\pm 0.4}\lsun$\\
$E(B-V)^{a}$ & 0.175\\
Ellipticity & $0.62^{+0.17}_{-0.21}$\\
Position angle (E of N) & $-24\pm16\deg$\\
major axis $r_{h}$ & $0\farcm57^{+0.32}_{-0.18}$\\
& $9.1^{+5.9}_{-3.4}\pc$\\
azimuthally averaged $r_{h}$ & $0\farcm44^{+0.18}_{-0.14}$\\
& $7.1^{+3.5}_{-2.4}\pc$\\
$^a$ from \citet{schlegel98}
\end{tabular}
\end{table}

\section{Discussion}
We have presented the discovery of a new stellar system, SMASH~1, that was found in the data from the SMASH survey. The system is located at a distance of $\sim57\kpc$; it is both faint ($L_V=10^{2.3\pm 0.4}\lsun$) and compact ($r_h=9.1^{+5.9}_{-3.4}\pc$).

\begin{figure*}
\begin{center}
\includegraphics[width=0.57\hsize,angle=270]{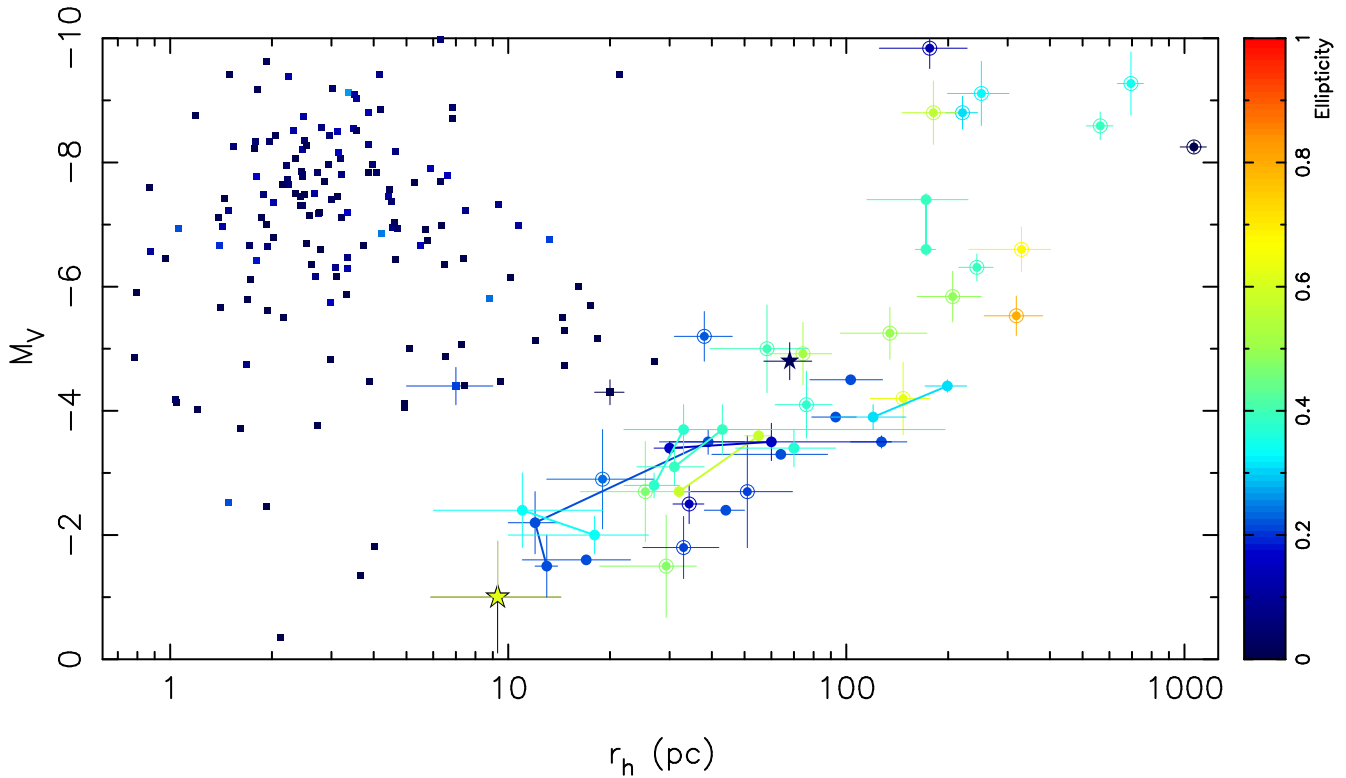}
\caption{\label{rh_Mv}Distribution of Milky Way satellites in the size--luminosity plane, where $r_h$ represents the major axis half-light radius. The colors encode the ellipticity of the objects, as represented by the color bar on the right. Globular clusters are shown as square and large squares with error bars represent the recently discovered Crater/Laevens~1 and Laevens~3 \citep{belokurov14,laevens14,laevens15b}. Milky Way dwarf galaxies are shown as circled dots (data from \citealt{irwin95}, \citealt{martin08b}, \citealt{belokurov09}, \citealt{belokurov10}, \citealt{laevens15b}, \citealt{torrealba16}). DECam-enabled discoveries are shown as large dots \citep[][for this latter reference, only high-confidence candidates are considered]{bechtol15, kim15a,kim15b,kim16,koposov15,drlica-wagner15}, except for the two SMASH discoveries that are shown as stars (SMASH~1, which appears in yellow at the bottom of the plot, and Hya~II, \citealt{martin15}). Lines link different literature measurements for a given system.}
\end{center}
\end{figure*}

SMASH~1 is located $11.3\deg$ away from the LMC in projection. Combined with its heliocentric distance, this places it $\sim13\kpc$ away from the Cloud (and $\sim20\kpc$ away from the SMC). Although quite distant from the LMC, SMASH~1 is located at a similar distance from the LMC than NGC~1841. More importantly, SMASH~1 is well within the tidal radius of the LMC, determined by \citet{vandermarel14} to be at least $16\kpc$ and potentially as large as $22\pm5\kpc$. Altogether, we conclude that SMASH~1 is likely a satellite globular cluster of the LMC, even though its location in the MW satellite size-luminosity plane is a little ambiguous (Figure~\ref{rh_Mv}, but see below).

As an old and metal-poor stellar system, it is natural to expect that SMASH~1 belongs to the LMC halo component. Alternatively, its properties are also spatially compatible with it being a (distant) LMC disk cluster. At the location of SMASH~1, the disk model of \citet{vandermarel14} has a heliocentric distance of $\sim55\kpc$, close to that of the new satellite. SMASH~1 would then be located at a disk radius of $11.3\kpc$ in the same model, or $\sim8$ scale-lengths \citep[e.g.,][]{vandermarel02}. Even though its low metallicity could be in tension with that of the bulk of LMC disk stellar populations, note that stars with such low metallicity exist in the outer disk of the LMC \citep{carrera11}. Velocities are necessary to determine whether or not SMASH~1 is a satellite of the LMC and if it traces its disk, as do other old LMC clusters \citep[e.g.,][]{olszewski91}.

Either way, the distance of SMASH~1 relative to the LMC raises interesting questions about the survivability of this satellite. The dynamical tidal radius, $r_t$, of a system can be calculated as follows (equation 7 of \citealt{innanen83}):

\begin{equation}
r_t \simeq 0.5 \left(\frac{M_\mathrm{SMASH1}}{M_{\mathrm{LMC}}(R_\mathrm{SMASH1})}\right)^{1/3} R_\mathrm{SMASH1},
\end{equation}

\noindent with $M_\mathrm{SMASH1}$ the stellar mass of the cluster, $R_\mathrm{SMASH1}$ its distance from the LMC, and $M_{\mathrm{LMC}}(R_\mathrm{SMASH1})$ the mass of the LMC enclosed within $R_\mathrm{SMASH1}$. With an assumed $M/L\sim2$ for SMASH~1's old and metal-poor stellar population \citep[e.g.,][]{pryor93} and $M_{\mathrm{LMC}}(R_\mathrm{SMASH1})>1.7\times10^{10}\msun$ (as measured by \citealt{vandermarel14} at a distance of $8.7\kpc$), we calculate $r_t\lta19\pc$ for SMASH~1. The tidal radius of the system is therefore much smaller than its spatial extent of $\sim4r_h=36\pc$ and SMASH~1 must be undergoing tidal disruption (see Figure~\ref{map}, in which the tidal radius is represented by the dashed blue circle).

That the cluster is undergoing tidal disruption could very well explain the high ellipticity we measure ($\epsilon=0.62^{+0.17}_{-0.21}$). In addition, as can be seen in Figure~\ref{map}, the position angle of the major axis points straight towards the LMC's centroid ($-24\pm16\deg$ vs. $-26\deg$), as expected if the system's stars are being pulled away from the system by the tidal forces of the LMC. Finally, if SMASH~1 really is disrupting, its original size would have been smaller, shifting the system into a part of the $r_h$--$M_V$ plane that hosts the faint globular clusters Koposov~1 and 2 \citep{koposov07} or AM-4 \citep{inman87}. Note, also, that the system undergoing tidal disruption would also explain its exceptionally large ellipticity compared to other LMC clusters \citep{bica08} and why its major axis points towards the LMC, something that is also not common among other LMC clusters. SMASH~1 has exceptional properties that can all be explained if it is assumed to be a satellite cluster of the LMC. However, this does not prove that this hypothesis is valid and, as pointed out above, a measure of the systemic velocity of the system is necessary to confirm its association to the Cloud.

Alternatively, SMASH~1 could be a dark-matter-dominated object. In this case, the tidal radius would be much larger than that calculated from the stellar component alone and the system would be shielded from tides. SMASH~1 would then join the cohort of recent faint satellites found around Magellanic system and that are mainly thought to be very faint dwarf galaxies \citep{drlica-wagner15,bechtol15,kim15a,koposov15,martin15}. However,the new system resides in a part of the $r_h$--$M_V$ plane in which no system has had its nature confirmed so far ($r_h\sim10\pc$, $-2<M_V<0$, see Figure~\ref{rh_Mv}; Kim~2, \citealt{kim15a}; DES~1, \citealt{luque16}; Mu\~noz~1, \citealt{munoz12b}; Eri~III, \citealt{bechtol15}). It will be arduous to measure the velocity dispersion of the system from its very few likely RGB stars or the more numerous but faint main sequence stars, and, from there, infer its dynamical mass. Therefore, the best hope to discriminate the nature of SMASH~1 is likely to come from a constraint on its spectroscopic metallicity dispersion.

In conclusion, we favor the scenario of a tidally disrupting globular cluster as it naturally explains the large elongation of the system pointing towards the LMC and the fact that it is very faint but rather extended for a cluster, especially at this distance from the LMC. SMASH~1 would then be one of only a few known disrupting globular clusters and the first such object to be discovered around the LMC.

What remains to be explained is the outstanding timing of witnessing SMASH~1 in the final throes of its tidal demise around the LMC, especially since it hosts an old stellar population and therefore needs to have survived around the Cloud(s) for a long time. 
In this context, an interesting solution is presented by \citet{carpintero13}, who showed that a significant fraction of outer LMC clusters could have been captured from the SMC. From a more benign orbit around the LMC or the SMC, the interaction between the two Clouds could well have sent SMASH~1 on the destructive orbit we observe it on today.

\acknowledgments

BCC acknowledges the support of the Discovery Grant DP150100862. Based on observations at Cerro Tololo Inter-American Observatory, National Optical Astronomy Observatory (NOAO Prop. ID: 2013B-0440; PI: Nidever), which is operated by the Association of Universities for Research in Astronomy (AURA) under a cooperative agreement with the National Science Foundation. This project used data obtained with the Dark Energy Camera (DECam), which was constructed by the Dark Energy Survey (DES) collaborating institutions: Argonne National Lab, University of California Santa Cruz, University of Cambridge, Centro de Investigaciones Energeticas, Medioambientales y Tecnologicas-Madrid, University of Chicago, University College London, DES-Brazil consortium, University of Edinburgh, ETH-Zurich, University of Illinois at Urbana-Champaign, Institut de Ciencies de l'Espai, Institut de Fisica d'Altes Energies, Lawrence Berkeley National Lab, Ludwig-Maximilians Universit\"at, University of Michigan, National Optical Astronomy Observatory, University of Nottingham, Ohio State University, University of Pennsylvania, University of Portsmouth, SLAC National Lab, Stanford University, University of Sussex, and Texas A\&M University. Funding for DES, including DECam, has been provided by the U.S. Department of Energy, National Science Foundation, Ministry of Education and Science (Spain), Science and Technology Facilities Council (UK), Higher Education Funding Council (England), National Center for Supercomputing Applications, Kavli Institute for Cosmological Physics, Financiadora de Estudos e Projetos, Funda\c{c}\~ao Carlos Chagas Filho de Amparo a Pesquisa, Conselho Nacional de Desenvolvimento Cient'fico e Tecnol\'ogico and the Minist\'erio da Ci\^encia e Tecnologia (Brazil), the German Research Foundation-sponsored cluster of excellence ``Origin and Structure of the Universe'' and the DES collaborating institutions. This research was made possible through the use of the AAVSO Photometric All-Sky Survey (APASS), funded by the Robert Martin Ayers Sciences Fund.


\end{document}